# Improved DC Bus Voltage Control of Single-Phase Grid-Connected Voltage Source Converters for Minimising Bus Capacitance and Line Current Harmonics

Sakda Somkun[1*], Viboon Chunkag[2]

[1] School of Renewable Energy Technology, Naresuan University, Phitsanulok, Thailand
[2] Department of Electrical and Computer Engineering, King Mongkut's University of Technology North Bangkok, Bangkok, Thailand
[*] sakdaso@nu.ac.th

**Abstract:** There is a trade-off between transient performance and line current distortion of the DC bus voltage control of single-phase grid connected voltage source converters. This paper presents an improved DC bus voltage control scheme of such converters using a proportional integral controller in series with a first order low pass filter. The extended symmetrical tuning method was adopted in the design of a regulator parameter, which greatly reduced the oscillating component at the double line frequency. The proposed control methodology allowed the loop bandwidth to increase without distorting the line current. Consequently, the DC bus voltage fluctuation or DC bus capacitance was reduced with a shorter settling time during a step load, compared with the conventional scheme. The proposed voltage scheme was found to be robust to the grid voltage variation and was less susceptible to the line voltage harmonics. Simulation and experimental results of a 1.5 kVA PWM rectifier verified the proposed methodology.

## 1. Introduction

Single-phase voltage source converters (VSCs) are normally used to interface the utility grid with renewable energy sources [1-3], energy storage devices , railway traction systems [4, 5], variable speed drives [6], and uninterruptable power supply systems [7]. In recent years an emerging application of such converters in the on-board battery chargers of electric vehicles (EVs) and plug-in hybrid electric vehicles (PHEVs) has drawn much attention from research communities because the so called *vehicle to grid* (*V2G*) operations can be used to supply active and/or reactive power back to the grid when needed [8, 9]. There are two common configurations of the single-phase grid-connected VSCs, i.e. single stage and multistage topologies. For the single stage topology, the DC bus of the VSC is directly connected to the energy sources or storage devices where the VSC supplies active power to the grid until the DC bus voltage reduces to the maximum power point [1, 2]. For the commonly used multistage topology as shown in Fig. 1, the DC bus voltage is connected to the second stage converter, which is usually one or more DC-DC converters [3, 7, 8, 10, 11] or three-phase DC-AC converters [4-6] for delivering power to the load and vice versa . The control system of such VSCs is usually of the cascade configuration. The fast inner loop is for regulation of the grid current, where a stationary-frame proportional-resonant (PR) controller [12, 13] or a rotating-frame proportional-integral (PI) controller  is successfully adopted to achieve a zero steady state line current error [14, 15]. For the outer DC bus voltage control loop, a PI regulator is normally employed to control the DC bus voltage greater than the peak value of the line voltage [4, 8, 10, 11]. The output signal of the DC bus voltage controller is then multiplied with a unity sinusoidal waveform obtained from the phase-locked loop (PLL), that is in phase with the grid voltage, to form the reference for the grid current control loop. This reference signal is only used for regulating the active power of the VSC, and it can be added with another sinusoidal signal orthogonal to the line voltage to inject reactive power to the grid for ancillary services in modern distributed generation systems [8]. The pulsating instantaneous power of the single-phase system intrinsically produces a double line frequency ripple in the DC bus voltage [16]. This ripple voltage problem does not arise in the counter-part three-phase VSCs due to the symmetry of the instantaneous power in the balanced three-phase system [13, 17].

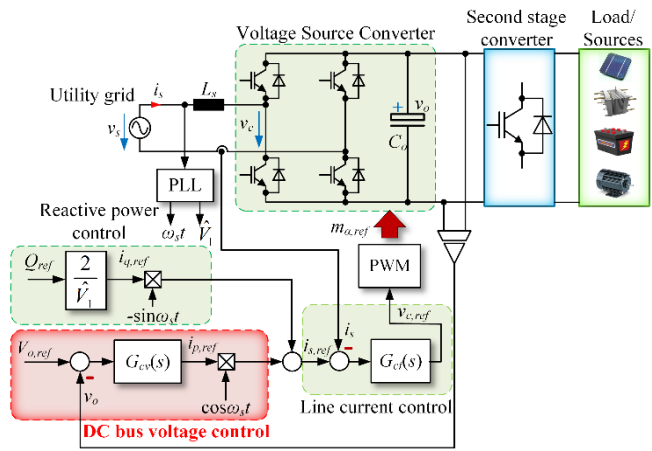

*Fig. 1.* DC bus voltage control system of two-stage single-phase grid-connected renewable energy system

The presence of the double line frequency ripple in the DC bus voltage control loop causes the third harmonic in the line current and also an unintentional reactive power injection [11]. Therefore, the DC bus voltage control loop is normally tuned at a bandwidth much lower than the double line frequency. This in turn causes a large bus voltage fluctuation during the transient condition. Bulky aluminium electrolytic capacitors are used to limit the DC bus voltage fluctuation, which increase the converter physical volume. Furthermore, reliability problems in in the aluminium electrolytic capacitors have driven research efforts to reduce the bus capacitance down to the range of more reliable film capacitors. The oscillating instantaneous power can be decoupled from the average power by using active ripple cancellation circuits [18-20]. With this approach, only the average power flows to the DC bus, which results in a lower bus capacitance and a higher loop bandwidth [20]. However, this requires extra semiconductor switches, passive components and additional control schemes, which are not suitable for low power or low cost applications. On the other hand, the ripple voltage can be permitted to some extent so as to reduce the bus capacitance and the ripple component is minimised in the DC bus voltage control loop by means of a notch filter [10], ripple estimators [21, 22], or average DC bus voltage measuring schemes [23-25]. These approaches allow a faster response and lower voltage fluctuation during the transient state. However, the notch filtering method needs to be adapted with the grid frequency and the dynamic of the notch filter must be taken into account when designing the PI regulator [10]. The ripple voltage estimating schemes require the peak value and phase angle of the line current to be extracted by a signal differentiator [21] or a PLL, which requires half of the line period to complete [22]. The average DC bus detecting schemes reported in [24] and [23] also perform every half of the line period. In [24], the fast DC bus voltage regulation was obtained by the feed forward output power and the DC bus voltage sensing gave a clean reference signal for the line current control loop. In [23], the DC bus voltage was not kept constant and it was changed with the bus power by means of an adaptive droop technique. So, the slow sampling rate in the DC bus voltage measurement was insignificant. A nonlinear observer was used to detect the peak value of the DC bus voltage which gave a fast DC bus regulating scheme [25]. However, the average DC bus voltage decreased when the bus power increased and vice versa. This was due to the fact that the peak value of the DC bus voltage was the controlled variable.

The aforementioned problems have indicated the need for a DC bus voltage control scheme that reduces the capacitance with an acceptable line current distortion. This paper proposes another control approach of the DC bus voltage of the single-phase grid-tied VSCs by using a low pass filter (LF) connected in series with a PI regulator. The extended symmetrical optimum method is used to design the parameters of the LF and PI controllers to obtain the desired bus voltage transient fluctuation and line current total harmonic distortion (THDi). Simulation and experimental results of a 1.5 kVA pulse width modulated (PWM) rectifier validate the proposed methodology, which are compared with the conventional control scheme.

## 2. System description and modelling

In this paper, the VSC under study is a PWM rectifier as depicted in Fig. 1. The line voltage is given by

$$v_s(t) = \hat{V}_1 \cos\omega_s t. \tag{1}$$

where $\omega_s$ is the grid frequency and $\hat{V}_1$ is the peak voltage value of the fundamental component. The desired line current is regulated by the current control loop as close as possible to

$$i_s(t) = \hat{I}_1 \cos(\omega_s t + \varphi) \tag{2a}$$

$$i_s(t) = i_p \cos\omega_s t - i_q \sin\omega_s t \tag{2b}$$

where $\hat{I}_1$ is the peak fundamental current, and $i_p = \hat{I}_1 \cos\varphi$ and $i_q = \hat{I}_1 \sin\varphi$ are the active and reactive power producing currents. The power balance in the VSC is written as

$$p_s(t) - p_{Rs}(t) - p_{Ls}(t) = p_{Co}(t) + p_o(t) \qquad (3)$$

where $p_s(t)$ is the instantaneous grid power, $p_{Ls}(t)$ is the instantaneous power in the line reactor, $p_{Rs}(t)$ is the instantaneous power in the winding resistance of the line reactor summed with losses in the power switches, $p_{Co}(t)$ is the instantaneous power at the DC bus capacitor, and $p_o(t)$ is the instantaneous output power. Neglecting losses in the winding resistance of the inductor and in the power switches, the DC bus voltage $v_o(t)$ is given by

$$v_s(t)i_s(t) - i_s(t)L_s \frac{di_s(t)}{dt} = v_o(t)C_o \frac{dv_o(t)}{dt} - v_o(t)i_o(t). \qquad (4)$$

In general, the oscillating power from the grid is mostly absorbed by the DC bus capacitor and the average DC bus voltage $V_o(t)$ is controlled close to the reference bus voltage $V_{o,ref}$. Thus, equation (4) can be written as

$$v_s(t)i_s(t) - i_s(t)L_s \frac{di_s(t)}{dt} \cong V_{o,ref} C_o \frac{dv_o(t)}{dt} - V_o(t)I_o(t). \qquad (5)$$

The dynamic of the average DC bus voltage is determined by taking an integration of (5) over a half of the line voltage period, which results in

$$V_{o,ref} C_o \frac{dV_o(t)}{dt} \cong \frac{\hat{V}_s}{2} i_p - P_o(t). \qquad (6)$$

Substituting (6) from (5) yields the bus ripple voltage $\tilde{v}_o(t)$ which is given by

$$C_o V_{o,ref} \frac{d\tilde{v}_o(t)}{dt} \cong \tilde{p}_s(t) - \tilde{p}_{Ls}(t). \qquad (7)$$

At the steady state condition, the oscillating components are written as

$$C_o V_{o,ref} \frac{d\tilde{v}_o(t)}{dt} \cong \frac{\hat{V}_1 \hat{I}_1}{2} \cos(2\omega_s t + \phi) + \frac{\omega_s L_s \hat{I}_1^2}{2} \sin(2\omega_s t + 2\phi). \qquad (8)$$

The instantaneous power $\tilde{p}_{Ls}(t)$ across the line inductor is comparatively small and can be neglected. Thus, $\tilde{v}_o(t)$ can be then approximated as

$$\tilde{v}_o(t) \cong \frac{1}{C_o V_{o,ref}} \int \frac{\hat{V}_1 \hat{I}_1}{2} \cos(2\omega_s t + \phi) dt \cong \frac{\hat{V}_1 \hat{I}_1}{4\omega_s C_o V_{o,ref}} \sin(2\omega_s t + \phi). \qquad (9)$$

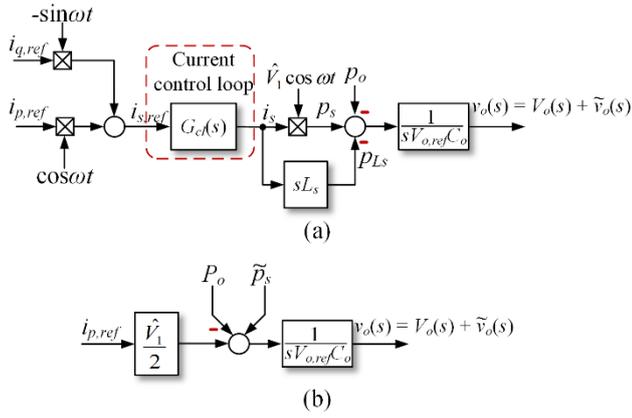

*Fig. 2.* DC bus voltage block diagrams
a Detailed block diagram
b Simplified block diagram

Fig. 2*a* explicates the detailed DC bus voltage block diagram derived from (5). The model is a time varying nonlinear system since it contains multiplication terms with the sinusoidal signals. The current control loop $G_{cl}(s)$ is normally much faster than the voltage control loop. Therefore, it can be considered as a unity gain. By separating the average components from the oscillating terms as given in (6) and (7), the average DC bus voltage block diagram can be simplified as illustrated in Fig. 2*b*, where the instantaneous power across the line reactor is neglected. In this model, the grid pulsating power $\tilde{p}_s(t)$ is considered as a disturbance causing the bus ripple voltage, whereas the average power $P_o(s)$ creates the average bus voltage fluctuation during the transient condition. It is worth mentioning that the simplified model in Fig. 2*b* is accurate for the control loop bandwidth below twice that the grid frequency which is typical for the DC bus voltage control.

## 3. Conventional DC bus voltage control scheme

The standard PI controller is used in the conventional DC bus voltage control scheme of the single-phase VSC as depicted in Fig. 3. In this control structure, the simplified DC bus voltage block diagram shown in Fig. 2 (b) is used and the high bandwidth line current control loop is approximated as a unity gain.

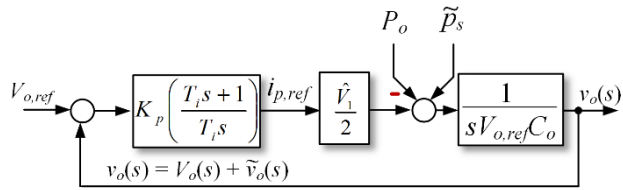

*Fig. 3.* Conventional DC bus voltage control scheme

The closed loop transfer function of the average DC bus voltage becomes

$$\frac{V_o(s)}{V_{o,ref}(s)} = \frac{T_i s + 1}{\left(2V_{o,ref} C_o T_i / \hat{V}_1 K_p\right) s^2 + T_i s + 1}. \tag{10}$$

The PI regulator is designed by comparison with the standard second order system given by

$$\frac{V_o(s)}{V_{o,ref}(s)} = G_{vl}(s) = \frac{(2\xi/\omega_n)s + 1}{s^2/\omega_n^2 + 2\xi s/\omega_n + 1} \tag{11}$$

where $\omega_n$ is the natural frequency and $\xi$ the damping ratio of the closed loop system. Thus, the parameters $K_p$ and $T_i$ are determined from

$$K_p = 2\xi\omega_n \cdot \frac{2V_{o,ref} C_o}{\hat{V}_1} \tag{12}$$

and

$$T_i = \frac{2\xi}{\omega_n}. \tag{13}$$

The bus reference voltage $V_{o,ref}$ is normally set at a constant value greater than the peak value of the line voltage. Therefore, it is more relevant to evaluate the recovery of $V_o$ due to a step change in the bus power $P_o$, which is expressed as

$$\frac{V_o(s)}{P_o(s)} = G_{dp}(s) = -\frac{1}{V_{o,ref} C_o} \cdot \frac{s/\omega_n^2}{s^2/\omega_n^2 + 2\xi s/\omega_n + 1}. \tag{14}$$

Another design criterion is the line current harmonics. This is caused by the presence of the oscillating component in the reference current $i_{p,ref}$, which can be determined from the attenuation of the ripple power $\tilde{p}_s$ as follows

$$\frac{i_{p,ref}(s)}{\tilde{p}_s(s)} = \frac{2}{\hat{V}_1} \cdot \frac{(2\xi/\omega_n)s + 1}{s^2/\omega_n^2 + 2\xi s/\omega_n + 1} = \frac{2}{\hat{V}_1} \cdot G_{vl}(s). \tag{15}$$

From (15), the oscillating component of $i_{p,ref}$ at the double line frequency $2\omega_s$ is written as

$$\tilde{i}_{p,ref}(j2\omega_s) = \frac{2}{\hat{V}_s} \cdot G_{vl}(j2\omega_s) \cdot \tilde{p}_s(j2\omega_s). \tag{16}$$

Thus, $\tilde{i}_{p,ref}(t)$ can be expressed as follows

$$\tilde{i}_{p,ref}(t) = \frac{2}{\hat{V}_s} \cdot |G_{vl}(j2\omega_s)| \cdot \frac{\hat{V}_1 \hat{I}_1}{2} \cos(2\omega_s t + \phi + \phi_{Gvl})$$
$$\tilde{i}_{p,ref}(t) = |G_{vl}(j2\omega_s)| \hat{I}_1 \cos(2\omega_s t + \phi + \phi_{Gvl}) \qquad (17)$$

where $\phi_{Gvl}$ is the phase angle of $G_{vl}(j2\omega_s)$. It is noted that the PI regulator produces the average $i_{p,ref}$ at the value of $\hat{I}_s \cos\phi$ corresponding to the bus average power and associated losses in the power switches and line reactor. Therefore, $\tilde{i}_{p,ref}(t)$ will be translated to the line current as follows

$$i'_s(t) = \tilde{i}_{p,ref}(t) \cdot \cos\omega_s t = |G_{vl}(j2\omega_s)| \hat{I}_1 \cos(2\omega_s t + \phi + \phi_{Gvl}) \cdot \cos\omega_s t$$
$$i'_s(t) = \frac{1}{2} |G_{vl}(j2\omega_s)| \hat{I}_1 [\cos(\omega_s t + \phi + \phi_{Gvl}) + \cos(3\omega_s t + \phi + \phi_{Gvl})] \qquad (18)$$

Equation (18) indicates that an additional reactive power of $|G_{vl}(j2\omega_s)| V_1 I_1 \sin\phi_{Gvl}$ is injected into the grid. More importantly, the ripple power creates the third harmonic in the line current for which the percentage value is given by [11]

$$\%I_3 = 50 |G_{vl}(j2\omega_s)|. \qquad (19)$$

Thus, THD$_i$ can be approximated as [11]

$$\%\text{THD}_i \approx 50 |G_{vl}(j2\omega_s)|. \qquad (20)$$

Equation (18) and (19) illustrate that the DC bus control loop $G_{vl}(s)$ should be tuned to have $|G_{vl}(j2\omega_s)|$ as low as possible to minimise the additional amount of the third harmonic and reactive power that is fed into the grid. The line current harmonics do not depend on the bus capacitance as it can be observed from the characteristics of $G_{vl}(s)$ in (11). This is achieved by selecting $\omega_n$ further below $2\omega_s$. However, a lower bandwidth creates a large bus voltage fluctuation during a power jump. Larger bus capacitance should be used to limit such voltage fluctuation as indicated in (14). For the 50 Hz system $\omega_s = 100\pi$ rad/s, if $G_{vl}(s)$ is designed at $\omega_n = 20\pi$ rad/s and $\xi = 1/\sqrt{2}$, $|G_{vl}(j2\omega_s)| = -17$ dB or $|G_{vl}(j2\omega_s)| = 0.141$. This will create THD$_i \approx 7\%$. Equation (15) indicates that the attenuation rate of $G_{vl}(s)$ after $\omega_n$ is -20 dB/Decade. Thus, a control scheme providing a sharper attenuation rate would reduce the line current distortion so that the bandwidth can be extended. This is the aim of this paper as presented in the next section.

## 4. Improved DC bus voltage control scheme

Fig. 4 depicts the proposed improved DC bus voltage control scheme where a first-order low-pass filter (LF) is added in series with the PI regulator. The LF is used to help the control loop attenuate the ripple power further down to the rate of -40 dB/Decade. The key idea of this proposed methodology is the adoption of the extended *symmetrical optimum* (SO) method [26] in selecting the LF time constant $T_f$ and the PI parameters $K_p$ and $T_i$.

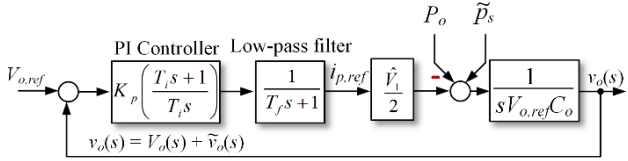

*Fig. 4.* DC bus voltage control system of two-stage single-phase grid-connected renewable energy system

From Fig. 4, the open loop transfer function of the DC bus voltage is written as

$$G_{ov}(s) = K_p \left( \frac{T_i s + 1}{T_i s} \right) \frac{1}{T_f s + 1} \left( \frac{\hat{V}_1}{2V_{o,ref} C_o} \right) \frac{1}{s}. \quad (21)$$

According the extended SO method, the parameter $K_p$ and $T_i$ are selected with the constant $\beta$ as follows

$$\left. \begin{array}{l} T_i = \beta T_f \\ K_p = \dfrac{1}{\beta^{1/2} T_f} \cdot \left( \dfrac{2V_{o,ref} C_o}{\hat{V}_1} \right) \end{array} \right\}. \quad (22)$$

This results in the maximum phase margin $\theta_{max}$ symmetrically around the cross-over frequency $\omega_o$ as illustrated in Fig. 5. The values of $\theta_{max}$ and $\omega_o$ are given by

$$\theta_{max} = \tan^{-1} \left( \frac{\beta - 1}{2\beta^{1/2}} \right) \quad (23)$$

$$\omega_o = \frac{1}{\beta^{1/2} T_f}. \quad (24)$$

The recommended values of $\beta$ are from 4 to 16, which relates to $\theta_{max}$ of 36° to 60°. With this setting, the closed-loop transfer function of the DC bus voltage becomes

$$\frac{V_o(s)}{V_{o,ref}(s)} = G_{vl}(s) = \frac{\beta T_f s + 1}{\beta \beta^{1/2} T_f^3 s^3 + \beta \beta^{1/2} T_f^2 s^2 + \beta T_f s + 1} \quad (25)$$

or

$$\frac{V_o(s)}{V_{o,ref}(s)} = G_{vl}(s) = \frac{\beta^{1/2} s / \omega_n + 1}{s^3 / \omega_n^3 + \beta^{1/2} s^2 / \omega_n^2 + \beta^{1/2} s / \omega_n + 1} \quad (26)$$

where

$$\omega_n = \omega_o = \frac{1}{\beta^{1/2}T_f}.\qquad(27)$$

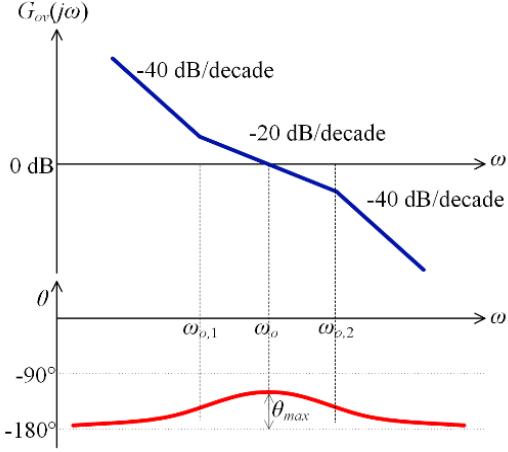

*Fig. 5.  Open loop Bode plot of the improved DC bus voltage control*

Similar to the conventional control scheme, the attenuation ratio to the ripple power is given by

$$\frac{i_{p,ref}(s)}{\tilde{p}_s(s)} = \frac{2}{\hat{V}_1}\cdot G_{vl}(s) = \frac{2}{\hat{V}_1}\cdot\frac{\beta^{1/2}s/\omega_n + 1}{s^3/\omega_n^3 + \beta^{1/2}s^2/\omega_n^2 + \beta^{1/2}s/\omega_n + 1}.\qquad(28)$$

The transfer function $G_{vl}(s)$ of the improved control strategy is the third order system with a zero, which provides an attenuation rate of -40 dB/Decade for the frequency greater than $\omega_n$. The disturbance rejection due to a change in the average bus power is also written as

$$\frac{V_o(s)}{P_o(s)} = G_{dp}(s) = -\frac{1}{V_{o,ref}C_o}\cdot\frac{\beta^{1/2}s/\omega_n^2(s/\beta^{1/2}\omega_n + 1)}{s^3/\omega_n^3 + \beta^{1/2}s^2/\omega_n^2 + \beta^{1/2}s/\omega_n + 1}\qquad(29)$$

which indicates that the bus voltage $V_o(t)$ will return to the reference value after a power step applied to the DC bus. The design procedure of this improved control scheme can be summarised as follows

1) Chose a value of $\beta$ for the desired phase margin using (23)
2) Calculate $\omega_n$ to create the desired magnitude $|G_{vl}(j2\omega_s)|$ from (26)
3) Calculate the time constant $T_f$ from (27)
4) Determine $K_p$ and $T_i$ from (22).

## 5. Performance evaluation

The conventional and improved DC bus control strategies are compared at the phase margins $\theta_{max}$ of 45° and 60°. For the conventional control scheme, this results in $\xi = 0.42$ at $\theta_{max} = 45°$ and $\xi = 0.61$ at $\theta_{max} = 60°$. According to (23), $\beta = 5.83$ at $\theta_{max} = 45°$ and $\beta = 13.9$ at $\theta_{max} = 60°$ for the improved control method. Fig. 6 shows the percentage of the third harmonic in the line current of the conventional and improved control strategies calculated from (19). If the third harmonic current is permitted at 2% of the fundamental current, the proposed control strategy has a higher bandwidth than the conventional control method at both phase margins. A lower phase margin reduces the third harmonic current, but it also decreases robustness of the control loop. Fig. 7 compares the closed-loop Bode plots of the improved and conventional control schemes at $\theta_{max} = 45°$. It shows that a steep attenuation ratio of -40 dB/Decade allows the bandwidth of the improved control method to increase to 12.93 Hz, whereas the bandwidth of the conventional scheme must be reduced to 4.85 Hz to limit the third harmonic within 2% of the fundamental component

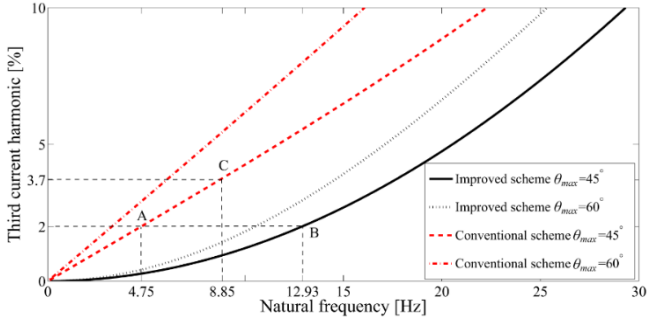

*Fig. 6.  Third harmonic content in the line current of the conventional and improved DC bus control schemes designed at the phase margins of 45° and 60°*

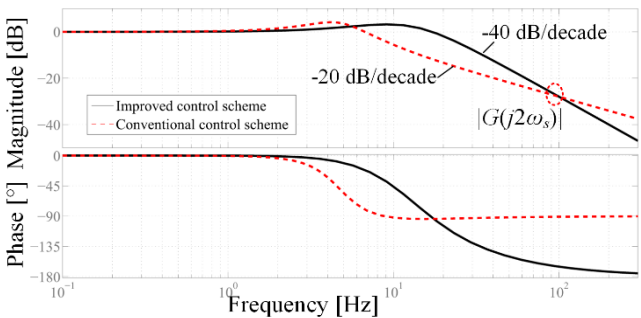

*Fig. 7.  Closed-loop Bode plots of the line current of the conventional and improved DC bus control schemes designed at $|G_{vl}(j2\omega_s)| = 0.04$ and the same phase margin of 45°*

Another factor to evaluate the performance of the control schemes is the maximum voltage fluctuation $\Delta V_{o,max}$ when the bus power $P_o$ changes, which is determined from

$$\Delta V_{o,max} = \max|e_v(t)| \tag{30}$$

where $e_v(t)$ is the bus voltage error obtained from the inverse Laplace transform of (14) and (29) as below

$$e_v(t) = \mathcal{L}^{-1}\left\{G_{dp}(s)\frac{P_o}{s}\right\}. \tag{31}$$

Fig. 8 compares the maximum bus voltage fluctuation of the conventional and improved control methods with the VSC parameters listed in Table 1 at $\theta_{max} = 45°$ and $\theta_{max} = 60°$. The improved control scheme exhibits a larger voltage fluctuation for the entire range of the loop bandwidth. This is due to its intrinsic third order system. However, under the same constraint of 2% third harmonic content, the improved method tuned at 12.93 Hz and $\theta_{max} = 45°$ (point B) has a voltage fluctuation of 24.1 V, whilst that of the conventional scheme designed at 4.75 Hz and $\theta_{max} = 45°$ (point A) is 45 V. This implies that the improved control method allows the bus capacitance to reduce to 589 µF, if the voltage fluctuation is kept the same as that of the conventional scheme. The increase of the phase margin reduces the voltage fluctuation at the same bandwidth for both control schemes with a trade-off being a higher third harmonic current. If the third harmonic current is constrained at 2%, the bandwidth of the conventional and improved schemes at $\theta_{max} = 60°$ must be decreased to 3.26 Hz and 10.54 Hz, which will eventually increase the voltage fluctuation. It is noted that the maximum voltage fluctuation is limited by the peak of the line voltage, which is $V_{o,ref} - \sqrt{2}V_s = 74.73$ V. From the control point of view, the phase margin of 60° is too high and is unnecessary. The phase margin of 45° is then chosen in the time domain simulation and experimental implementation of the VSC.

**Table 1** Parameters for the VSC under study

| Parameters | Values |
|---|---|
| Nominal RMS line voltage, $V_s$ | 230 V |
| Nominal grid frequency, $\omega_s$ | 100π rad/s |
| Reference DC bus voltage, $V_{o,ref}$ | 400 V |
| Maximum active power, $P_o$ | 1 kW |
| Maximum reactive power, $Q$ | 1 kVar |
| DC bus capacitance, $C_o$ | 1.1 mF |
| Line reactor, $L_s$ | 8.2 mH |
| Line reactor winding resistance, $R_s$ | 0.68 Ω |
| Switching frequency, $f_{sw}$ | 10 kHz |

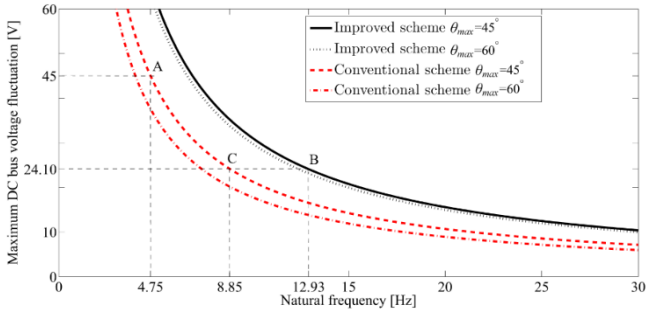

***Fig. 8.*** *Maximum DC bus voltage fluctuation of the conventional and improved control strategies due to a power jump of 1 kW for $V_{o,ref} = 400$ V and $C_o = 1.1$ mF at the phase margins of 45° and 60°*

The integral time absolute error (ITAE) criterion is also used to characterise how the DC control methods recover to the reference voltage after a power jump. The ITAE criterion is defined by

$$ITAE = \int_0^T t|e_v(t)|dt \qquad (32)$$

where $T$ should be large enough for $e_v(t)$ to reach the steady state. In this work, $T = 5$ seconds is chosen. Fig. 9 displays the ITAE criteria of the conventional and improved control methods for $P_o = $ 1kW, $V_{o,ref} = 400$ V and $C_o = 1.1$ mF. Again, the improved control method has the ITAE criterion greater than that of the conventional method. However, at the 2% third harmonic current and $\theta_{max} = 45°$, the ITAE criterion of the conventional scheme at the bandwidth of 4.75 Hz (point A) is shown to be 0.35 V·s², which is about 15 times greater than that of the improved strategy at 12.93 Hz (point B).

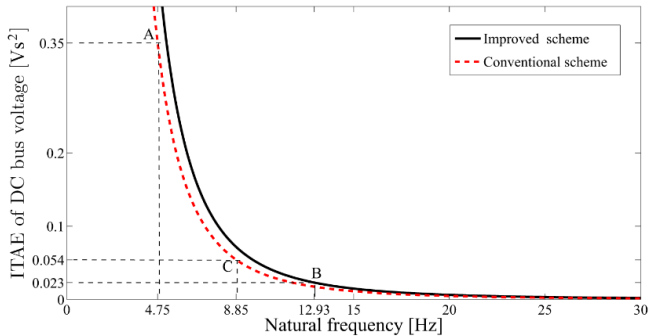

***Fig. 9.*** *ITAE criteria of the conventional and improved control strategies due to a power jump of 1 kW for $V_{o,ref} = 400$ V and $C_o = 1.1$ mF at the phase margins of 45° and 60°*

Another benefit of the SO tuning method is that it is well suited to a variable gain process. This is due to the fact that the maximum phase margin that occurs symmetrically around the cross-over frequency decreases slightly when the process gain varies from the nominal value. Observing from the simplified DC bus block diagram shown in Fig. 2b, the line voltage and bus capacitance both contribute to the process gain.

Fig. 10 illustrates the open loop Bode plots of the conventional and improved control strategies at the phase margins of 45° when the line voltage varies ± 30% from the nominal values. The phase margin of the conventional scheme is more sensitive to a drop in the line voltage than the proposed control method. When the line voltage increases to 30% of its nominal value, the phase margin of the proposed method reduces slightly from 45° to 44.4°, which is insignificant to the control loop stability. It can be concluded that the proposed DC bus voltage control methodology is more robust to a line voltage drop than the conventional method.

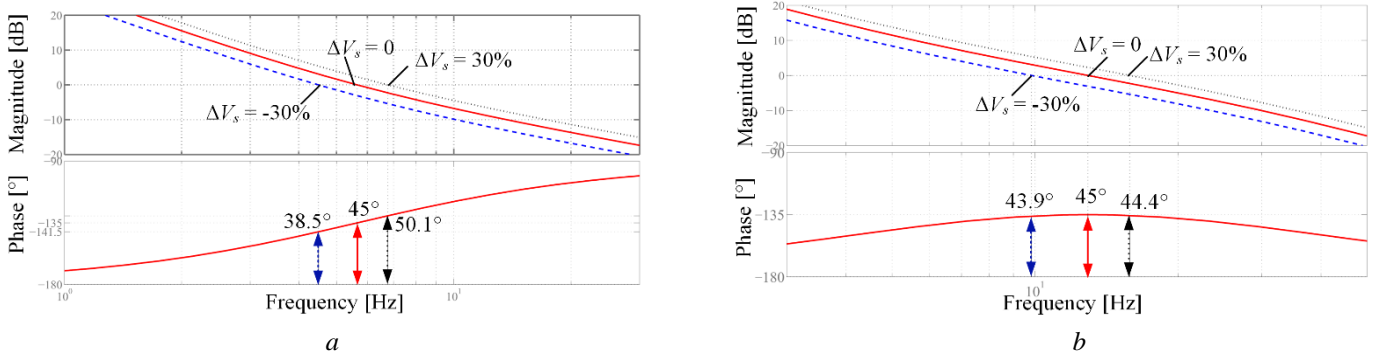

***Fig. 10.*** *Open loop Bode plot of the DC bus voltage control schemes at the phase margin of 45° subjected to line voltage variations*
*a* Conventional scheme
*b* Improved scheme

## 6. Simulations

A switched circuit model of the VSC with parameters is listed in Table 1. This model was implemented in the MATLAB/Simulink program, which was also compared with the simplified model as depicted in Fig. 2*b*. The unbalanced synchronous reference frame control strategy was adopted as the line current regulator, which was proven to provide a zero steady state error due to an infinite gain at the grid frequency [14]. The current loop was designed at a bandwidth of $2\pi \times 1,000$ rad/s, which is high enough to be considered as a unity gain seen by the voltage loop.

Four design examples of the DC bus voltage control as listed in Table 1 are compared. According to Fig. 6, the target third harmonic current is 2%, which results in $\omega_n = 2\pi \times 4.75$ rad/s for the conventional method (point A in Fig. 6, Fig. 8 and Fig. 9) and $\omega_n = 2\pi \times 12.93$ rad/s for the improved scheme (point B in Fig. 6, Fig. 8 and Fig. 9). Also, the conventional control strategy designed at $\omega_n = 2\pi \times 8.85$ rad/s and $\theta_{max} = 45°$ (point C in Fig. 6, Fig. 8 and Fig. 9) provided the same voltage fluctuation as the improved method. The conventional control strategy was therefore chosen as the third design example. The last design example is the improved scheme at $\omega_n = 2\pi \times 12.93$ rad/s with the bus capacitance reduced to $C_o = 0.68$ mF.

**Table 2** Design examples of the DC bus voltage control at the phase margin of 45°

| Design examples | $\omega_n$ [rad/s] | $C_o$ [mF] | $\Delta V_{o,max}$ [V] | $I_3$ [%] |
|---|---|---|---|---|
| 1. Conventional | $2\pi \times 4.75$ | 1.10 | 43.2 | 2 |
| 2. Improved | $2\pi \times 12.93$ | 1.10 | 23.1 | 2 |
| 3. Conventional | $2\pi \times 8.85$ | 1.10 | 23.1 | 3.76 |
| 4. Improved | $2\pi \times 12.93$ | 0.68 | 37.4 | 2 |

Simulation results of $V_o(t)$ and $i_{p,ref}(t)$ using design examples 1 to 4 are shown in Fig. 11. At the beginning, $V_o(t)$ rises from its initial value at the peak of the line voltage of 325 V to the reference value of 400 V without load. After $V_o(t)$ reaches the steady state, the DC bus is connected to a resistive load of 960 W (166.67 Ω), similar to that used in the experiments. Once, $V_o(t)$ is stabilised, the load is disconnected from the DC bus. The results from the simplified model (the dashed lines) are very close to those obtained from the switched circuit model that includes the line inductor and its winding resistance. This confirms the accuracy of the performance evaluation discussed in the previous section. The maximum voltage fluctuation is observed during the load disconnection because the suddenly removed load behaves like an ideal step load. The simulation results show that the maximum voltage fluctuations of the conventional and improved control schemes are very close to the analytical values indicated in Table 2. The settling time of the improved scheme shown in Fig. 11b is about 0.07 s, whereas the conventional method with 4.75 Hz bandwidth exhibited in Fig. 11a takes more than 0.3 s to recover to the reference bus voltage after the load is switched off. This confirms the calculated ITAE values shown in Fig. 9. The settling time of the conventional method with 8.85 Hz displayed in Fig. 11c is about 0.2 s, also longer than the improved control scheme even though the maximum voltage fluctuations of two case are similar. The transient response of the improved strategy with $C_o = 0.68$ mF indicated in Fig. 11d has a similar characteristic as that with $C_o = 1.1$ mF shown in Fig. 14, except it has a higher voltage fluctuation.

Fig. 12 shows the simulation results of the line voltage and current at $P_o = 960$ W of the design examples 1 to 4. The line currents are also compared with its fundamental component $i_{s1}(t)$ (dashed lines), which account for the output power and the copper loss $I_1 R_s$. Table 3 summarises the harmonic components of the line currents indicated in Fig. 12, which shows a good agreement with the predicted third harmonic depicted in Fig. 6 for the phase margin of 45°. A large ripple component in $i_{p,ref}$ of the conventional scheme at 8.85 Hz displayed in Fig. 11c results in a greater line current distortion. The line current harmonics of the design examples 2 and 4 also confirm (28) that the bus capacitance does not have the impact on the double line frequency attenuation [11]. Therefore, the proposed control scheme allows the DC bus capacitance to decrease to the minimum acceptable bus voltage fluctuation without increasing the line current distortion.

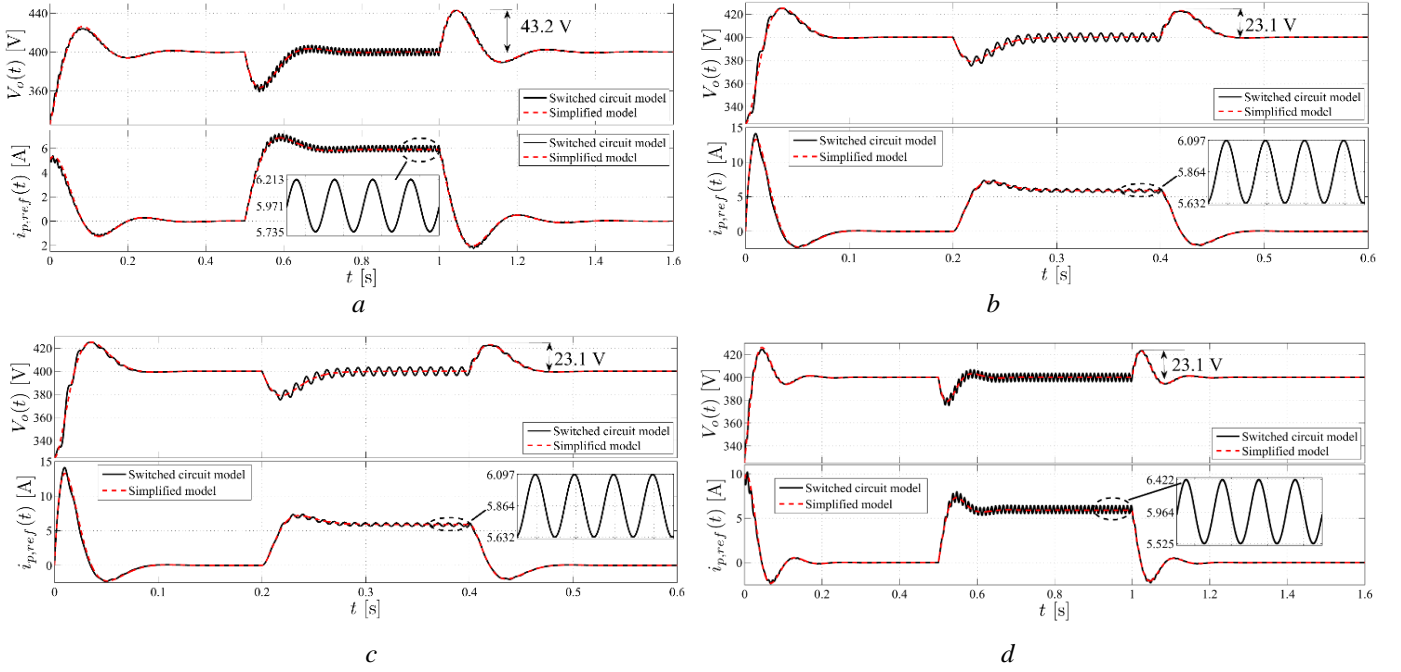

**Fig. 11.** *Simulated $V_o(t)$ and $i_{p,ref}(t)$ of the DC bus voltage control schemes when connecting a resistive load of 960 W, and vice versa*
*a* Conventional scheme at the bandwidth of 4.75 Hz (design example 1)
*b* Improved scheme at the bandwidth of 12.93 Hz (design example 2)
*c* Conventional scheme at the bandwidth of 8.85 Hz (design example 3)
*d* Improved scheme at the bandwidth of 12.93 Hz and $C_0 = 0.68$ mF (design example 4)

Another minor point that can be observed from the simulation are the differences in the average values of $i_{p,ref}$ ($I_{p,ref}$) exhibited in the insets of Fig. 11a to Fig. 11d. The values of $I_{p,ref}$ of the improved schemes in Fig. 11b are close to that in Fig. 11d (5.684 A and 5.685 A). $I_{p,ref}$ of the conventional methods is equal to 5.971 A for the bandwidth of 4.75 Hz, as shown in Fig. 11a, with $I_{p,ref}$ equal to 5.964 A for the bandwidth of 8.85 Hz (Fig. 11c). This is caused by an additional reactive power of $|G_{vl}(j2\omega_s)|V_1 I_1 \sin\phi_{Gvl}$ being injected into the grid, as explained by (18). This creates extra losses in the line reactor winding and in the switching devices, which in turn requires increased active power. Observing from the Bode plots in Fig. 7, $\phi_{Gvl}$ at 100 Hz of the proposed method is about -170°, whilst $\phi_{Gvl}$ of the conventional strategy is close to -90°. This results in $I_{p,ref}$ of the conventional control method being greater than that of the proposed scheme and is even higher at a lower bandwidth as $\phi_{Gvl}$ at $2\omega_s$ is getting closer to -90°. This resultant increased loss seems to be small and negligible, but it can be of significance for higher power applications such as in railway traction systems [4, 5].

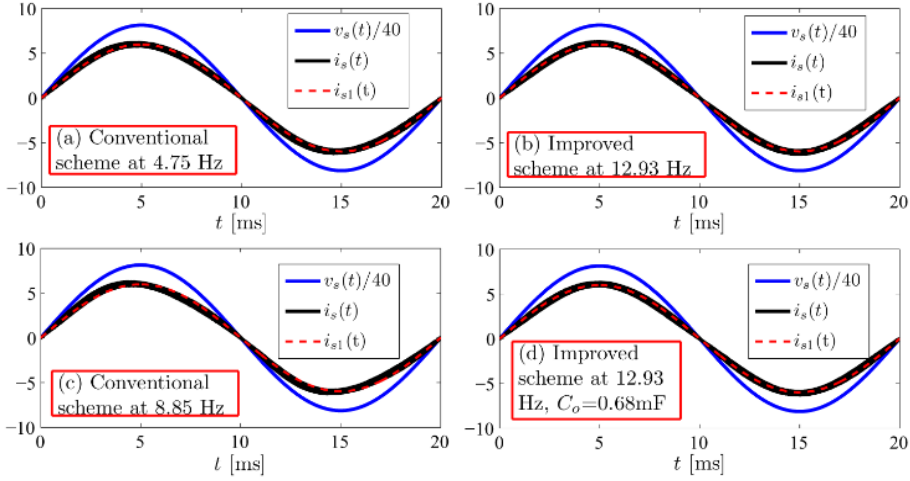

*Fig. 12.* Simulation results of $v_s(t)$ and $i_s(t)$ at $P_o = 960$ W of the design examples 1 to 4

**Table 3** Line current harmonic contents obtained from the simulation at $P_o = 960$ W

| Design examples | $\%I_3$ | $\%I_5$ | $\%I_7$ | $\%I_9$ |
|---|---|---|---|---|
| 1. Conventional 4.75 Hz, 1.1 mF | 2.07 | 0.03 | 0.01 | 0.02 |
| 2. Improved 12.93 Hz, 1.1 mF | 1.96 | 0.03 | 0.02 | 0.02 |
| 3. Conventional 8.85 Hz, 1.1 mF | 3.85 | 0.10 | 0.03 | 0.01 |
| 4. Improved 12.93 Hz, 0.68 mF | 1.94 | 0.02 | 0.01 | 0.02 |

## 7. Experimental validation

In this study, the VSC with the parameters given in Table 1 was constructed using components listed in Table 4. The control algorithm was implemented on a TMS320F28069 32-bit microcontroller from Texas Instruments. Measured signals $v_s(t), i_s(t)$ and $v_o(t)$ were synchronously sampled with the PWM at the frequency of 20 kKz, which was twice that of the switching frequency. The current loop was performed at the sampling rate of 20 kHz. The sampling frequency of the DC bus voltage control loop was 4 kHz. The dead time delay to prevent short circuits in each leg of the VSC was set at 250 ns. Grid synchronisation was obtained by a single-phase Park-based PLL as presented in [27]. Parameters of the PI regulator and the LF obtained from the simulations with proper scaling were directly used in the experiment. The reference current signal $i_{p,ref}(t)$ was transformed to a ±10 V signal using a DAC8802 14-bit digital to analogue converter.

**Table 4** Main components of the VSC under study

| Devices | Manufacturers and models |
|---|---|
| IGBT modules | IXYS MWI15-12A7 (1.2 kV, 17 A) |
| Gate drivers | 2 × Semikron SKHI-22B |
| Voltage sensors | 2 × LEM LV 25-NP (Hall effect) |
| Current sensor | LEM CAS 6-NP (Flux gate) |
| DC bus capacitor | 2 × EPCOS 2.2 mF 450V electrolytic (Series connection) KAMET 0.68 mF 550V electrolytic |

Fig. 13 displays the transient responses $V_o(t)$, $i_{p,ref}(t)$, $v_s(t)$ and $i_s(t)$ illustrated in design examples 1 to 4, which occurred when the resistive load of 960 W was disconnected to compare with the simulations shown in Fig. 11. The experimental waveforms agree with the simulations, except the experimental maximum voltage fluctuation of design example 1, shown in Fig. 13*a* was 38 V, which was slightly less than the simulated value of 43.1 V portrayed in Fig. 11*a*. This could be due to device tolerances of the bus capacitance and the load resistor, and the line RMS voltage during the experiment.

Fig. 14 shows the steady state waveforms $V_o(t)$, $i_{p,ref}(t)$, $v_s(t)$ and $i_s(t)$ of the design examples 1 to 4 when supplying output power of 960 W. The reference current $i_{p,ref}(t)$ of the conventional scheme with the bandwidth of 8.85 Hz (design example 3) has the highest peak to peak value, similar to the simulation. Moreover, $i_{p,ref}(t)$ of the conventional scheme with the bandwidth of 4.75 Hz (design example 1) has the greatest average value in response to the additional reactive power of $|G_{vl}(j2\omega_s)|V_1 I_1 \sin\phi_{Gvl}$ as analysed in the previous section. This can be observed from the difference in the phase angle of $i_{p,ref}(t)$ of the conventional scheme shown in Fig. 14*a* and that of the improved method depicted in Fig. 14*b*.

Table 6 compares the line current harmonic components of design examples 1 to 4, which were measured by a Metrel MD9272 power clamp meter. The background harmonics of the line voltage were measured before connecting the VSC to the grid, which are given in Table 7. The VSC was set to supply a capacitive reactive power of 1,000 Var to the grid for power quality ancillary services [8]. This indicates that the experimental line current harmonics agree with those obtained from the simulations with higher magnitudes due to the background harmonics of the grid voltage. However, the conventional method at 4.75 Hz and 8.85 Hz (design examples 1 and 3) seems to pick up the background harmonics more than the improved schemes do (design examples 2 and 4) when compared with the values from the simulation. This can be explained by using the Bode plots depicted in Fig. 7. In the conventional scheme, the magnitude of $G_{vl}(j\omega)$ for a frequency above $2\omega_s$ is greater than that of the improved scheme, which make it more susceptible to the line voltage harmonics.

The experimental results of the improved control schemes with different bus capacitance shown in Fig. 14*b* and Fig. 14*d*, also in Table 6 again confirm that only the control loop bandwidth has the impact on the line distortion. Furthermore, the results shown in Table 6 confirms (19) and (20) that the handling of reactive power does not alter the line current harmonic contents.

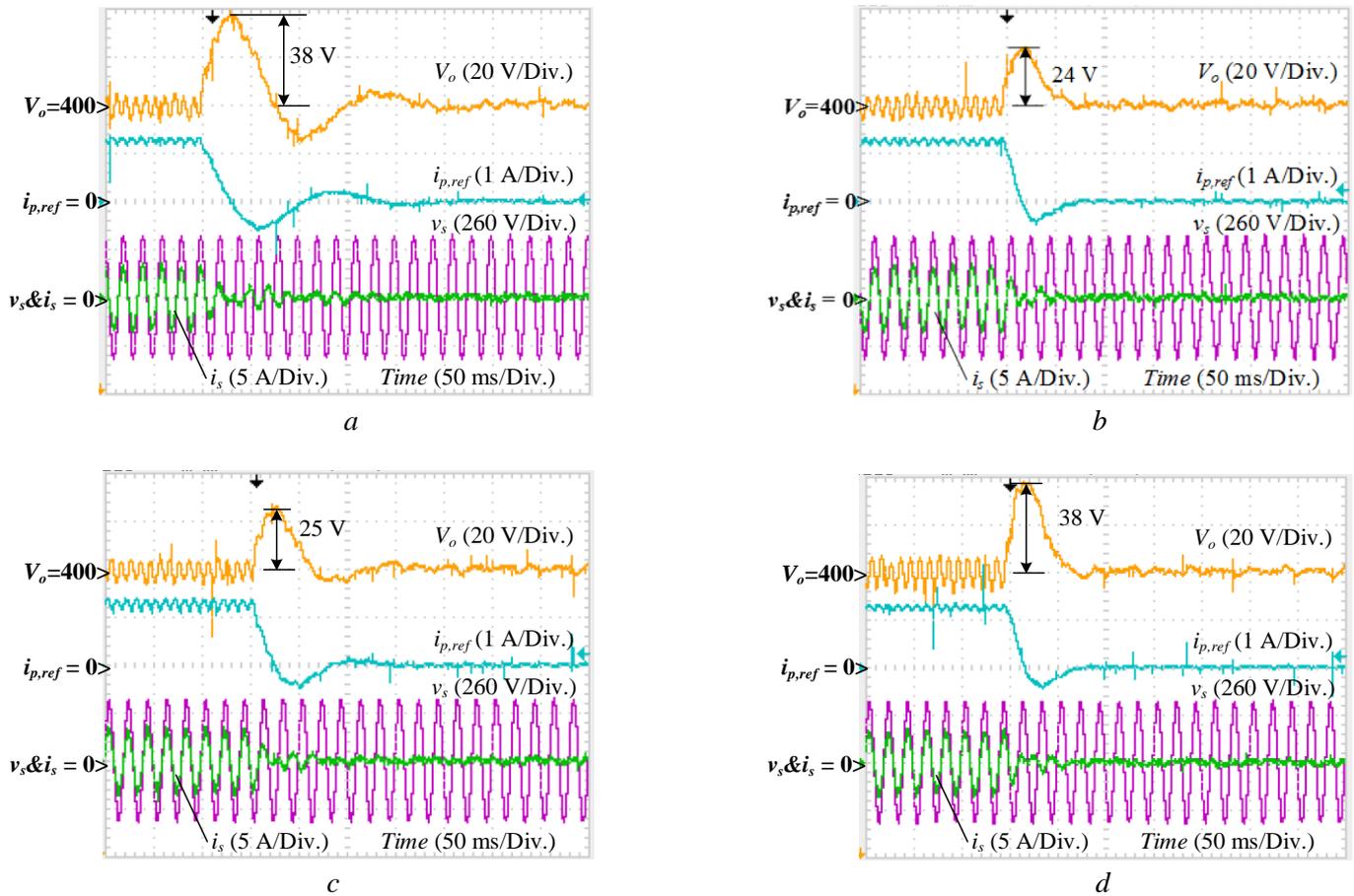

***Fig. 13.*** *Experimental results of* $V_o(t)$, $i_{p,ref}(t)$, $v_s(t)$ *and* $i_s(t)$ *of the DC bus voltage control schemes when disconnecting a resistive load of 960 W*
*a* Conventional scheme at the bandwidth of 4.75 Hz (design example 1)
*b* Improved scheme at the bandwidth of 12.93 Hz (design example 2)
*c* Conventional scheme at the bandwidth of 8.85 Hz (design example 3)
*d* Improved scheme at the bandwidth of 12.93 Hz and $C_0 = 0.68$ mF (design example 4)

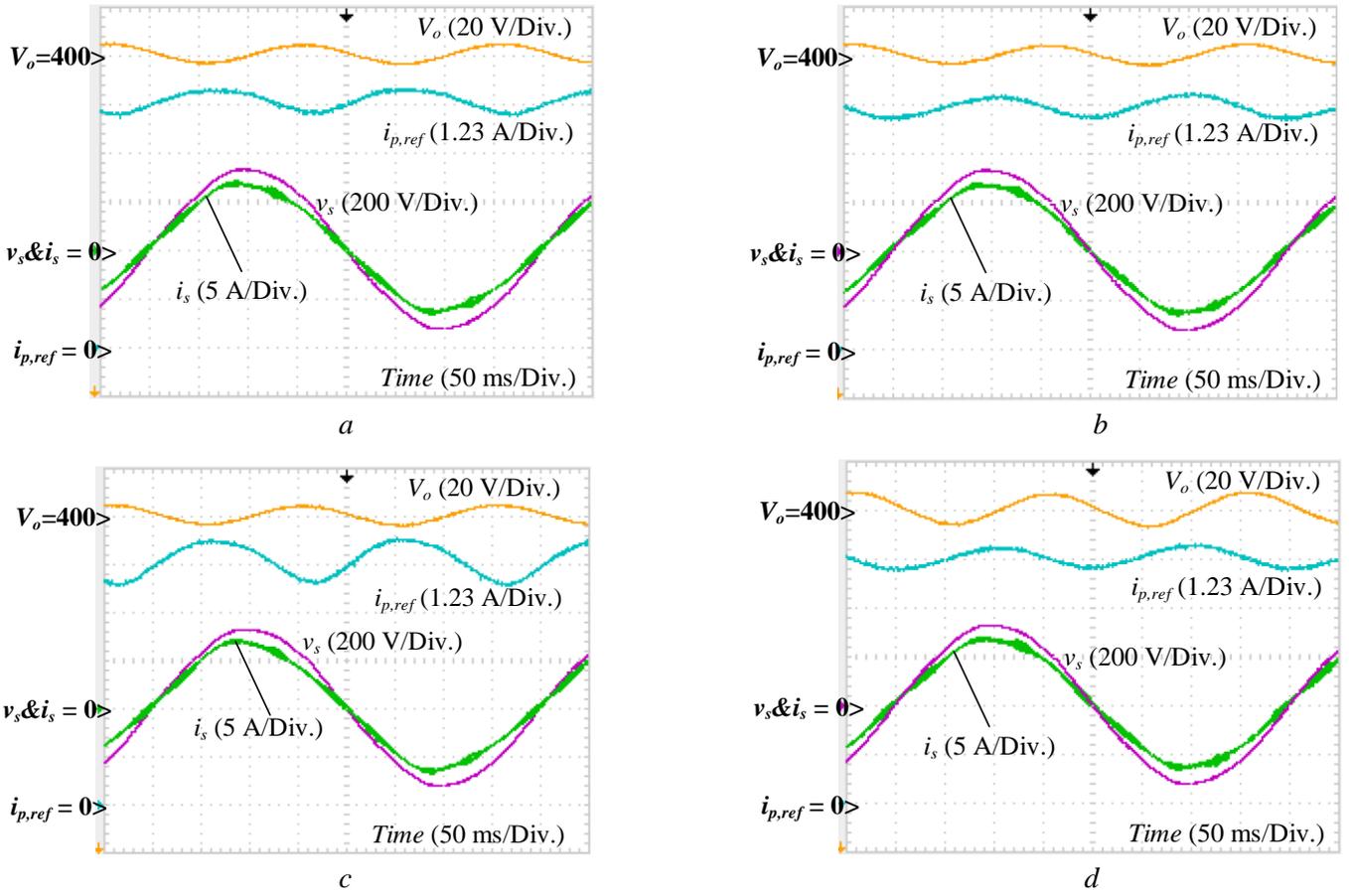

**Fig. 14.** *Steady state waveforms of* $V_o(t)$, $i_{p,ref}(t)$, $v_s(t)$ *and* $i_s(t)$ *at* $P_o = 960$ W
*a* Conventional scheme at the bandwidth of 4.75 Hz (design example 1)
*b* Improved scheme at the bandwidth of 12.93 Hz (design example 2)
*c* Conventional scheme at the bandwidth of 8.85 Hz (design example 3)
*d* Improved scheme at the bandwidth of 12.93 Hz and $C_0 = 0.68$ mF (design example 4)

**Table 6** Line current harmonic contents obtained from the experiments

| Design examples | $P_0$ [W] | $Q_{ref}$ [Var] | %$I_3$ | %$I_5$ | %$I_7$ | THD [%] |
|---|---|---|---|---|---|---|
| 1. Conventional 4.75 Hz, $C_0 = 1.1$ mF | 960 | 0 | 2.8 | 0.8 | 0 | 3.0 |
|  | 0 | 1,000 | 2.8 | 1.1 | 0 | 3.4 |
|  | 960 | 1,000 | 2.8 | 1.5 | 0 | 3.4 |
| 2. Improved 12.93 Hz $C_0 = 1.1$ mF | 960 | 0 | 1.9 | 1.3 | 0 | 2.7 |
|  | 0 | 1,000 | 1.9 | 1.1 | 0 | 2.4 |
|  | 960 | 1,000 | 1.9 | 1.1 | 0 | 2.3 |
| 3. Conventional 8.85 Hz $C_0 = 1.1$ mF | 960 | 0 | 4.6 | 2 | 0 | 5.2 |
|  | 0 | 1,000 | 4.6 | 1.9 | 0 | 5.1 |
|  | 960 | 1,000 | 4.9 | 1.5 | 0 | 5.2 |
| 4. Improved 12.93 Hz $C_0 = 0.68$ mF | 960 | 0 | 1.7 | 1.5 | 0 | 2.8 |
|  | 0 | 1,000 | 1.6 | 1.2 | 0 | 2.2 |
|  | 960 | 1,000 | 1.7 | 1.2 | 0 | 2.2 |

**Table 7** Line voltage harmonic components during the experiment

| $V_s$ [V] | %$V_3$ | %$V_5$ | %$V_7$ | %$V_9$ | THD [%] |
|---|---|---|---|---|---|
| 229.1 | 0.3 | 1.6 | 0.1 | 0.3 | 2.0 |

## 8. Discussion

We have demonstrated that the proposed DC bus control scheme consisting of a standard PI controller in series with a low pass filter and adopting a symmetrical optimum tuning methodology, has several advantages over the conventional control method with only a PI regulator. The advantages are:

1) Lower bus voltage fluctuation during a step load due to the increased bandwidth with the same line current harmonic contents.
2) The DC bus capacitor can be reduced with an acceptable voltage fluctuation.
3) Lower ITAE value during a step load due to a smaller settling time.
4) Robust to the line voltage drop due to the phase margin being maximised symmetrically around the nominal line voltage, which makes the proposed scheme suitable for low voltage ride through (LVRT) applications [1].
5) Smaller unintentional reactive power injection because the phase angle of the voltage control loop is approaching -180° at the frequency of $2\omega_s$.
6) Less susceptible to background harmonics of the grid voltage due to the steeper attenuation rate of -40 dB/Decade.
7) Easy to implement and the design of the regulator's parameters is straightforward.

Another approach to improve the DC bus voltage control can be done by using a notch filter in series with the PI regulator to suppress the $2\omega_s$ component [10]. In that paper, there was a bus voltage fluctuation of 20 V during a step load of 200 W with the bus capacitance of 112 µF. The DC bus voltage reference was 400 V and the line voltage was 240 V. Such voltage fluctuation is approximately inferred to the VSC used in this study as

$$20\,\text{V} \times \frac{960\,\text{W}}{200\,\text{W}} \times \frac{112\,\mu\text{F}}{1{,}100\,\mu\text{F}} = 9.8\,\text{V}$$

which is better than the improved control method proposed by us. However, the design of such a control strategy is complicated as the notch filter dynamic must be included in the loop transfer function. In addition, such a notch filter should be adaptable to the grid frequency, which would not, however, be convenient for analogue implementation. More importantly, the robustness of the notch filtering control scheme with a line voltage drop was not examined.

In comparison with the notch filtering method, the proposed scheme is easier to implement on both analogue and digital platforms. The proposed scheme can be constructed in an integrated circuit for a low cost single-phase VSC application. The controller parameters $K_p$, $T_i$ and $T_f$ are directly chosen from the desired bandwidth and the parameters of the VSC. It is also proven to be suitable for LVRT applications. Thus, it depends on design engineers' choice which approach is appropriate to their products. Also, the proposed DC bus voltage control scheme can be applicable to unidirectional power flow AC/DC converters [24, 28].

## 9. Conclusion

A DC bus voltage control strategy of the single-phase grid connected voltage source converters consisting of a proportional integral controller in series with a first order low pass filter together is proposed. The extended symmetrical optimum tuning method is used to design the parameters of the regulator and low pass filter. The proposed control scheme has a steeper attenuation rate of -40 dB/Decade to the double line frequency component, which increases the control loop bandwidth compared with the conventional control method. The proposed scheme reduces the bus voltage fluctuation or DC bus capacitance with the same line current harmonics. It was experimentally found that the proposed control scheme is less susceptible to the background harmonics of the grid voltage. As well, the proposed control methodology is more robust to the line voltage drop than the conventional strategy.

## 10. Acknowledgments

This work has been financially supported by the Thailand Research Fund (TRF) and Naresuan University, research grant no. TRG5780100. Many thanks to Mr. Roy Morien of the Naresuan University Language Centre for his editing assistance and advice on English expression in this document.